\documentclass[preprint,aps,pra]{revtex4-2}
\usepackage{graphicx}
\usepackage{siunitx}
\usepackage[hidelinks]{hyperref}
\usepackage{amsmath}
\usepackage[version=4]{mhchem}

\usepackage[range-phrase = \ --\ , range-units=single]{siunitx}
\DeclareSIUnit{\dpa}{dpa}
\DeclareSIUnit\angstrom{\text{\AA}}
\DeclareSIUnit\Angstrom{\text{\AA}}
\begin{document}

\title{Universal degradation of high-temperature superconductors due to impurity scattering: predicting the performance loss in fusion magnets}


\author{M.~Eisterer, A.~Bodenseher, R.~Unterrainer}
\affiliation{Atominstitut, TU Wien, Stadionallee 2, 1020 Vienna, Austria}
 
\begin{abstract}

Predicting the change of performance of superconductors under neutron radiation is indispensable for designing compact fusion devices. The favorable enhancement of the critical current caused by flux pinning is separated from the degrading effect of increased scattering of the charge carriers to derive a degradation function from the expected change of the superfluid density (reducing to Homes' law in the dirty limit) and the observed increase in flux creep. The degradation turned out to be a universal function of disorder, not depending on the particular tape nor the particle radiation: thermal and/or fast neutrons, as well as 1.2\,MeV protons. The universal behavior enables the analysis of changes in flux pinning corrected by the adverse enhancement of scattering. A more reliable prediction of the performance change of coated conductors in a fusion reactor based on proxies for neutrons is anticipated.   

\end{abstract}
\maketitle
\newpage


\section{Main}

Much effort has been devoted to an understanding of vortex pinning in high temperature superconductors (HTS) to optimize the critical current for applications.~\cite{Sado2016,Driscoll2021,Puig2024} However, with the idea of compact fusion devices promising a fast development of industrial fusion power plants, a new aspect of defect landscapes called for immediate attention. Particle radiation, inherent to nuclear fusion, adds defects to the highly optimized pinning structure of state-of-the-art HTS conductors and, after an initial increase of the tape´s performance, leads to a rapid degradation of the critical current ($I_\textrm{c}$),~\cite{Fischer2018,Unterrainer2022} posing severe limitations on the life time of a fusion magnet.~\cite{Torsello2022, Ledda2024} 
The physics of the increase of $I_\textrm{c}$ is in principle understood by the newly introduced pinning centers, but it is hard to predict because it depends on both, the defect structure of the pristine tape and the particle radiation itself. The degradation on the other hand, seems widely universal: The transition temperature, $T_\textrm{c}$, linearly decreases with fluence~\cite{Rullier2003,Fischer2018,Unterrainer2022,Unterrainer2024} at least within the range relevant for technical applications (change of $T_\textrm{c}$ by less than 20\,\%). The slope is independent of the tape but characteristic for each type of particles and their energies.~\cite{Unterrainer2024} Impurity scattering is pair breaking in the cuprate superconductors due to the $d$-wave symmetry of their order parameter.~\cite{Radtke1993} Since $T_\textrm{c}$ is predicted to decrease (initially) linearly with the scattering rate, too, pair breaking scattering is an obvious candidate for the degradation of the superconducting properties. The transition temperature itself becomes a suitable measure of impurity scattering with $D$ defined as the disorder resulting in a decrease of $T_\textrm{c}$ by 1\,K. The change of $T_\textrm{c}$ with fluence can be either determined experimentally, or by damage calculations for particle energy distributions not available in experiments (e.g. the neutron spectrum in fusion magnets).    
\par
\begin{figure}[t]
\begin{centering}
\includegraphics[width=0.8\linewidth]{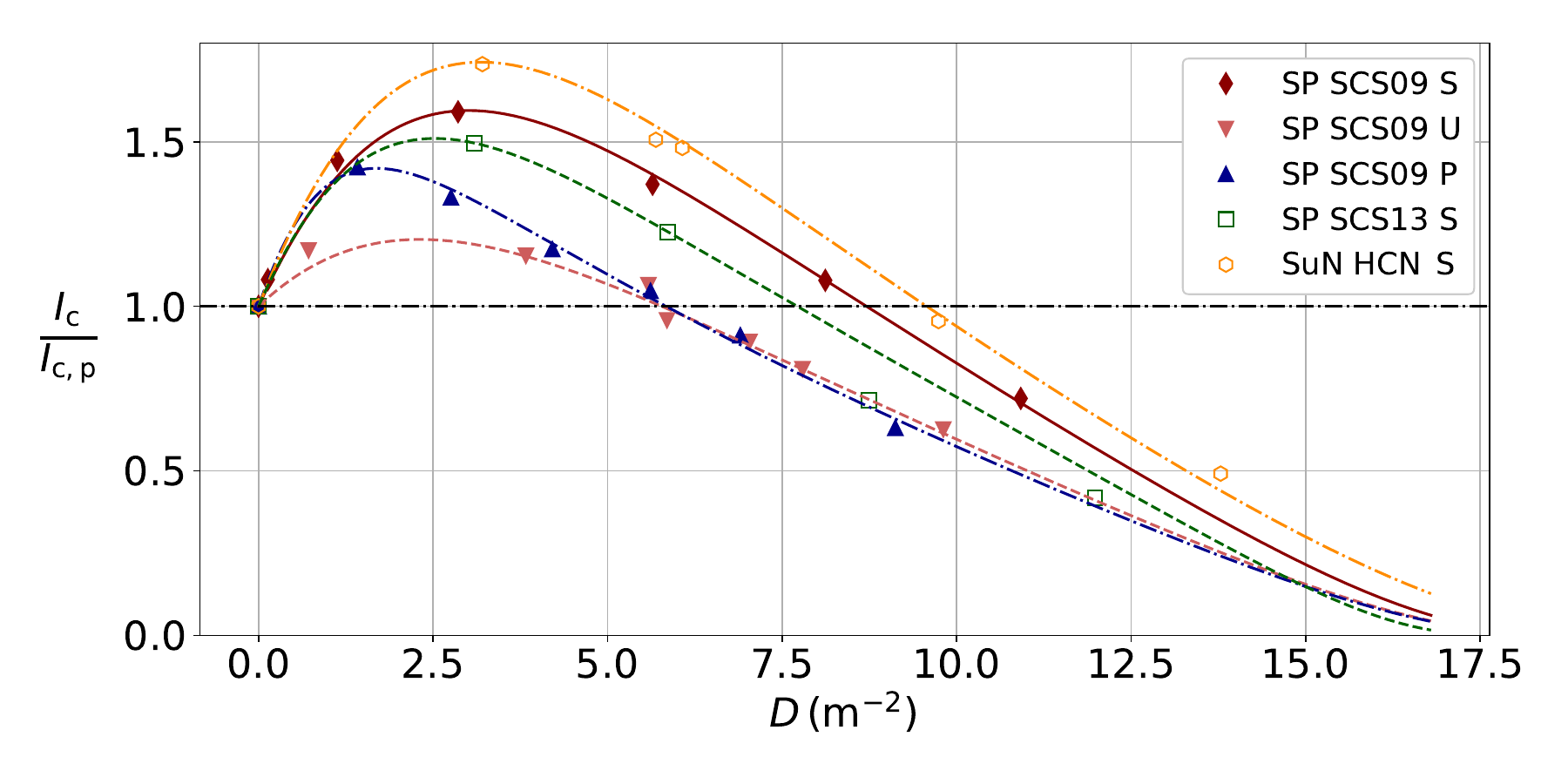} \caption{Relative change of the critical current by particle radiation at 15 T, 30 K. Different tapes where exposed to fast neutron (S), thermal and fast neutrons (U) and 1.2\,MeV protons (P). The line graphs refer to the modelling of the experimental data (symbols).}
\label{FigChangeIc}
\end{centering}
\end{figure}

\medskip
Fig.~\ref{FigChangeIc} compares the relative change of the (experimental) critical current (ratio of $I_\textrm{c}$ after and prior to fast neutron irradiation) as a function of disorder. Data obtained from the same tape but irradiated with different particles or energy spectra (solid symbols) and from two other tapes (open symbols) irradiated with fast neutrons are shown.  In this representation, the non-universality of the increasing part of the curve and the similar behavior of the decreasing part become evident.~\cite{Unterrainer2024} 

For separating the influence of enhanced pinning from the degradation by impurity scattering, the pinning efficiency, $\eta_\textrm{pin}$, is introduced: $J_\textrm{c}(B,T)=\eta_\textrm{pin}(B,T)J_\textrm{dp}(B=0,T)$. The depairing current density  $J_\textrm{dp}=\phi_0/3\sqrt{3}\pi\lambda^2\xi\mu_0$ is the absolute limit for loss free currents and depends on the fundamental superconducting material parameters, i.e., the magnetic penetration depth $\lambda$ and the superconducting coherence length $\xi$. They are both expected to change with disorder, hence, a decrease in $J_\textrm{dp}$ reflects the effect of scattering. The pinning efficiency on the other hand, cannot reach values much above 30\,\%~\cite{Arcos2005} and reflects the change of vortex pinning due to the introduced defects. 
\par\medskip
Finally, since any experiment measures  $J_\textrm{c,exp}$ at a certain electric field (a freely chosen value in direct transport measurements, commonly 1\,$\mu$V/cm), a flux creep correction, $A_\textrm{creep}$ has to be taken into account: $J_\textrm{c,exp}(B,T)=A_\textrm{creep}(B,T)J_\textrm{c}(B,T)$. $A_\textrm{creep}$ is always smaller than one since vortices are depinned almost immediately due to thermal activation processes at $J_\textrm{c}$ defined via the critical state model.~\cite{Anderson1961,Yeshurun1996} 
\par
\begin{figure}[ht!]
\begin{centering}
\includegraphics[width=0.8\linewidth]{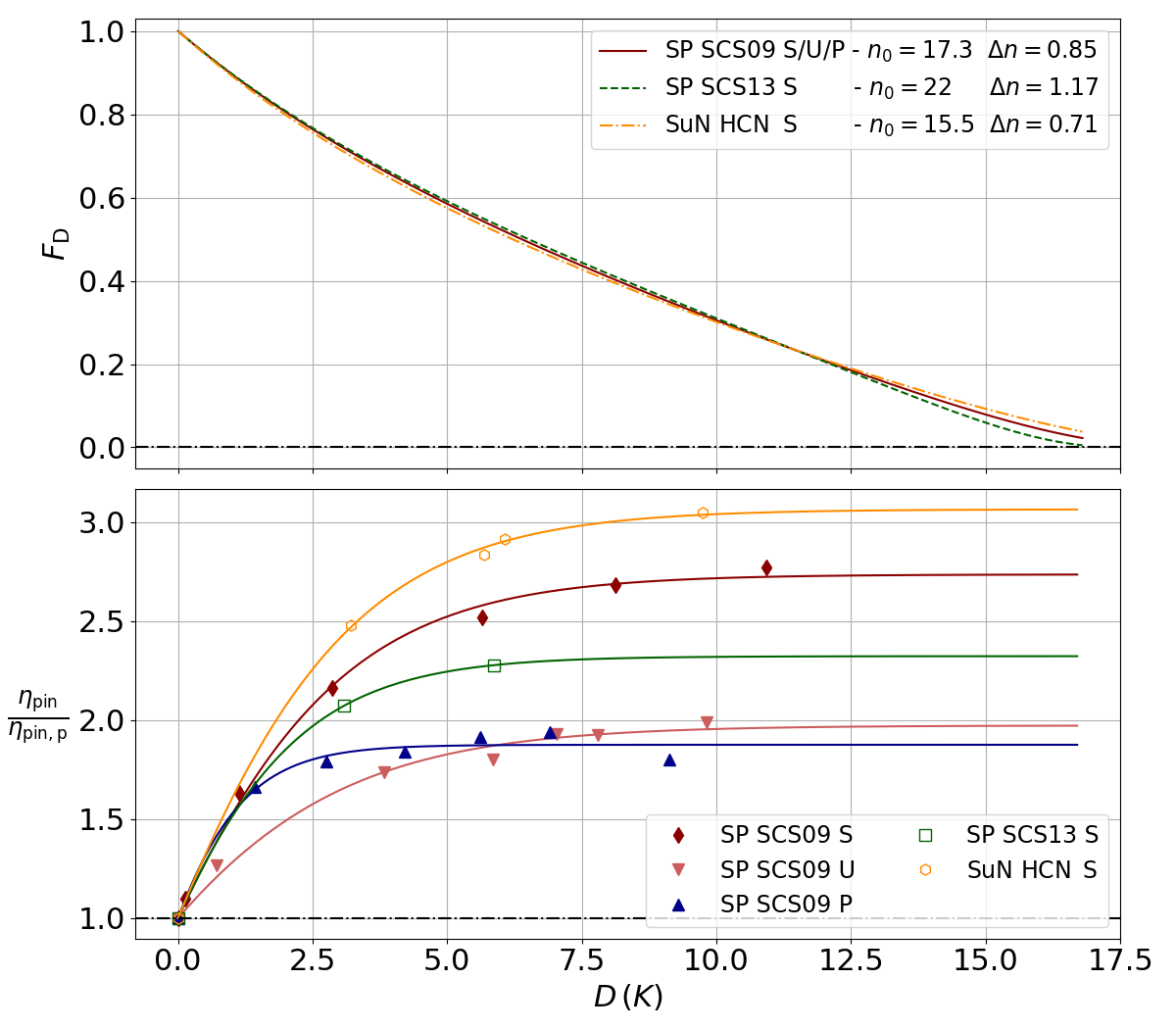}
\caption{Upper panel: The degradation function is nearly identically for all samples. $\Delta n = -\partial n / \partial D$. Lower panel: Change of pinning efficiency at \qty{15}{\tesla}, \qty{30}{\kelvin}. }
\label{fig:model}
\end{centering}
\end{figure}

\medskip
The influence of disorder on $J_\textrm{dp}$ and $A_\textrm{creep}$ is derived in the next section. The relative change of the critical current $I_\textrm{c,i}/I_\textrm{c,p}$ ($I_\textrm{c}$ always refers to the experimental value corresponding to   $J_\textrm{c,exp}$ to avoid excessive indexing) just becomes the product of the relative changes of the decreasing depairing current density, $A_\textrm{creep}$ (enhanced flux creep), and $\eta_\textrm{pin}$ (improved pinning). The adverse affect from the reduction in superfluid density and enhanced flux creep can be described by the degradation function $F_D(D)$ (Equ.\ref{EqDegrad}). The degradation function contains three parameters: $\alpha_\textrm{p}$ which refers to the ratio of clean limit coherence length and mean free path of the charge carriers in the pristine sample. It was set to three to render the theoretical and observed (in samples with high $T_\textrm{c}$) coherence length compatible. $F_D$ is insensitive to slight changes of $\alpha_\textrm{p}$. A large change of $\alpha_\textrm{p}$ would lead to a significant change of $T_\textrm{c}$, so it is expected to be similar in all coated conductors. The second parameter, the change in $n$-value with disorder, $\partial n/\partial D$, was assessed experimentally and found to vary weakly between different tapes when normalized to its pristine value, although the measurement error is high. Finally, the third parameter, $K_\rho=-16.5$, representing the relative change of the normal state resistivity ($\rho_\textrm{n}$) with changing $T_\textrm{c}$ was determined by neutron irradiation experiments on YBCO films, because it cannot be assessed easily in coated conductors due to the metallic stabilizing layers and substrate. It is expected to be universal since changes of both, $T_\textrm{c}$ and $\rho_\textrm{n}$, are based on the change in the scattering rate. Therefore, $F_D$ becomes nearly universal, which is confirmed by the similar degradation behavior~\cite{Unterrainer2024} demonstrated in Fig.~\ref{FigChangeIc} and the calculated degradation functions shown in the upper panel of Fig.~\ref{fig:model}.
\par\medskip
Knowing the detrimental effect of the radiation (or disorder in general) quantitatively, the change in pinning can be obtained by $\eta_\textrm{pin}/\eta_\textrm{pin,p}=I_\textrm{c}/I_\textrm{c,p}F_D$. The results for the data shown in Fig.~\ref{FigChangeIc} are displayed in the lower panel of Fig.~\ref{fig:model}. After an initial increase, the pinning efficiency saturates which is expected given the theoretic limit of $\eta_\textrm{pin}$. A decrease of $\eta_\textrm{pin}$ certainly occurs if too much of the superconductor is destroyed by the large defects, but for presented data the volume fraction of the large defects is too small to observe this effect.~\cite{Unterrainer2024}
\par\medskip
Defects efficient for pinning are not the same as those enhancing scattering, although in principle each defect is expected to contribute to both, pinning and scattering. Pinning is most efficient for defects with a radius comparable to the coherence length (or larger in the direction of the vortex core) while a shortening of the mean free path requires a high density of small defects. Since the distance between two large (nanometer sized) defects is much larger than the coherence length ($\lesssim 2$\,nm) they are not expected to add significantly to scattering, except in the clean limit. The same quantity of displaced atoms, if located in large versus small defects, does not equally contribute to pinning and pair breaking scattering.  For instance, neutrons with an energy above \qty{0.1}{\mega\electronvolt} create large defects relevant for pinning, called collision cascades, whereas small defects (relevant for scattering) are formed by lower energy neutrons or as a side effect of collision cascades. 

\newpage
The observed saturation behavior of $\eta_\textrm{pin}$ motivates to fit the change in pinning by $\eta_\textrm{pin}(D)=\eta_\textrm{pin}^\textrm{max}\tanh{((D+D_0)/D_\textrm{n})}$, with three free parameters: $\eta_\textrm{pin}^\textrm{max}$ giving the saturation value of $\eta_\textrm{pin}$, $D_0$ defining the pinning efficiency in the pristine conductor, and $D_\textrm{n}$ is related to the disorder where $\eta_\textrm{pin}$ saturates. The results of these fits (line graphs in Fig.~\ref{fig:model}, bottom panel) are used as the change of $\eta_\textrm{pin}$ due to the introduced pinning centers to describe the change of $I_\textrm{c}$ under irradiation. The agreement (line graphs in Fig.~\ref{fig:model}) with the experimental data is excellent. Pinning hardly changes on the degrading branch of the curves; thus, the behavior is given by $F_D$ only.
\par\medskip
The simultaneous interaction of different types of pinning centers with the vortex lattice is by means not trivial, hence out of the scope of this study. Nevertheless, data corrected by the degrading effect of disorder will certainly help to gain more insights into this difficult topic in the future.  
\par\medskip
In conclusion the derived degradation function is nearly universal, the minor influence of the change of $n$ with disorder can be obtained by easy to perform irradiation experiments (e.g. with protons). The change in pinning due to the introduction of pinning efficient defects is then directly obtained, independent of the simultaneous enhancement of scattering. This will foster the understanding of the interaction of different defects in improving the critical current and help finding suitable proxies for neutrons to reliably predict the change of performance of HTS conductors in fusion magnets without the need of expensive (cryogenic) neutron irradiation experiments. 



\section{Derivation of the degradation function}
The Gorkov-Goodman relation~\cite{Goodman1961} predicts the upper critical field, $H_\textrm{c2}=\Phi_0/2\pi\xi^2$, to increase linearly with resistivity from its clean limit value in $s$-wave superconductors, therefore, the coherence length changes as $\xi=\xi_0/\sqrt{1+\xi_0/l}$ with the mean free path of the charge carriers $l$ and the clean limit (BCS) coherence length $\xi_0$. This reduces to the widely used dirty limit expression  $\xi=\sqrt{\xi_0l}$ for $l<<\xi_0$. It follows from thermodynamics (Ginzburg-Landau theory) that the condensation energy is proportional to $(\lambda\xi)^{-2}$ and consequently $\lambda=\lambda_0\sqrt{1+\xi_0/l}$ since the condensation energy has to remain constant for non pair breaking scattering. $\lambda_0$ is the clean limit value of $\lambda$.
So far nothing that is obviously not applicable to unconventional superconductors has been used but to further proceed the BCS relations $\xi_0=\hbar v_\textrm{F}/\pi\Delta$ and $2\Delta=3.53(4.25)k_\textrm{B}T_\textrm{c}$ for $s$- and $d$-wave superconductors, respectively, are needed. However, the first relation is directly derived from the composition of the superconducting wave function  in $k$-space from states within the energy gap and hence a natural consequence of the width of the gap. Therefore, the coherence length (size of a cooper pair) becomes the inverse of the energy gap in real space. \par 
These are universal principles not relying to specifics of the underlying wave functions. The validity of the second relation for unconventional superconductors is less obvious, but experiments show that the ratio between $\Delta$ and  $T_\textrm{c}$ is not far off the BCS prediction.~\cite{Hüfner2008} Since pair breaking scattering reduces the transition temperature in $d$-wave superconductors it also increases $\xi_0$ as given by above relations. Expressing the mean free path of the charge carriers (density $n$) by the normal state resistivity $\rho_\textrm{n}=m_\textrm{eff} v_\textrm{F}/ne^2l$ and using the London expression for $\lambda_0^2=m_\textrm{eff}/\mu_0ne^2$  (obtained from electrodynamics) transforms $\lambda_0^2\xi_0/l$ to $K_\lambda\rho_\textrm{n}/T_\textrm{c}$ and the above relation for the increase of $\lambda$ by impurity scattering becomes
\begin{equation}
\lambda^2=\lambda_0^2+K_\lambda\frac{\rho_\textrm{n}}{T_\textrm{c}}.
\end{equation}
$K_\lambda=a\hbar/\mu_0k_\textrm{B}$ is free of any material dependent parameters, with $a$=0.18 (0.15) for $s$- ($d$-)wave superconductors. In the dirty limit, $\lambda_0^2$ on the right hand side can be neglected resulting in Homes' law~\cite{Homes2004} $\lambda^{-2}=K_\lambda^{-1}T_\textrm{c}/\rho_\textrm{n}$. This general trend was found empirically as an extension of the so-called Uemura plot originally observed in underdoped cuprates.~\cite{Uemura1989} Homes' law  was  derived from BCS theory in the dirty limit,~\cite{Kogan2013a} but the derivation given here better reveals the underlying physics and clarifies that it is in principal not a relation for the superfluid density itself ($\propto 1/\lambda^{2}$) but for $1/(\lambda^{2}-\lambda_0^{2})=1/(\lambda_0^2\cdot \xi_0/l)=:1/\alpha\lambda_0^2$.
\par
\begin{figure} \centering \includegraphics[, width = 0.6\columnwidth,clip]{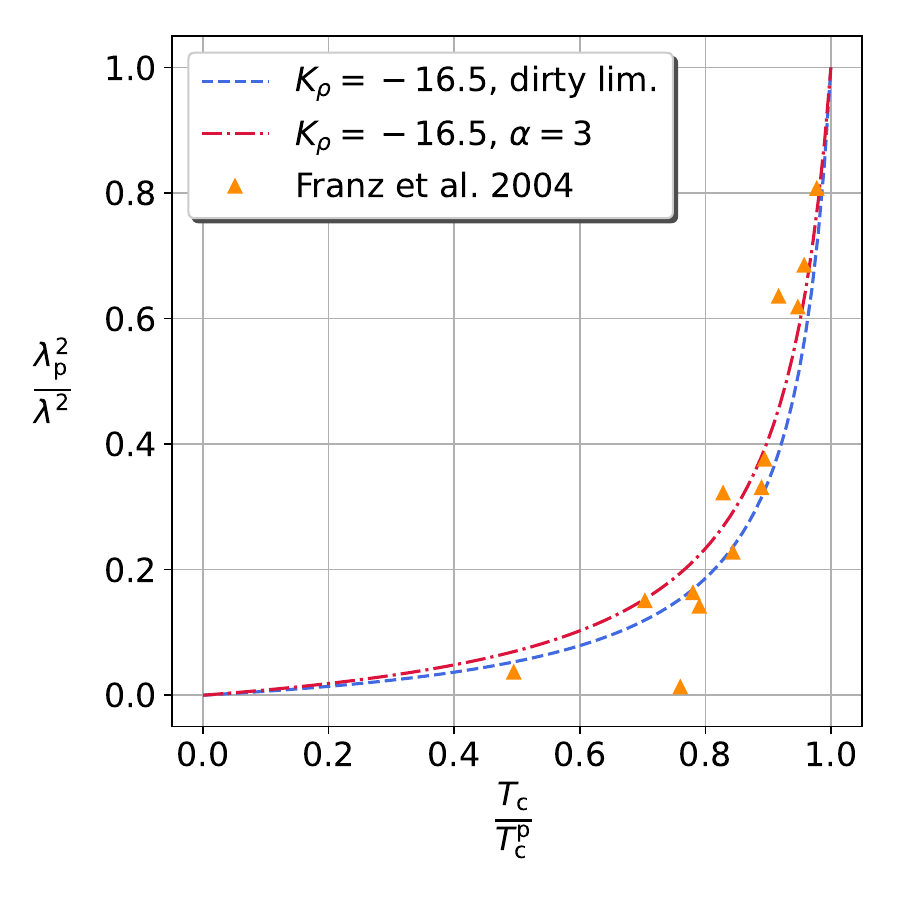} \caption{Relative change of superfluid density in YBCO thin films. Experimental data were extracted from Ref.~\cite{Franz1997}. 
Theoretical prediction was calculated with $K_\rho=-16.5$ and $\alpha_\textrm{p}=3$, or assuming dirty limit}\label{FigSuperfluid}
\end{figure}

\medskip
However, the dirty limit version has the advantage to abandon the a priori unknown $\lambda_0$, which is difficult to access experimentally and reduces the universality of Homes' law. Using the above BCS expressions for $\xi_0$, with $v_\textrm{F}=2.7\cdot10^5$m/s~\cite{zhou2003} and $T_\textrm{c}\simeq 90$\,K for YBCO results in $\xi_0\simeq 3.3$\,nm, about twice the typically reported values for $\xi$. This implies that $\alpha=\xi_0/l$ is about 3, making the dirty limit relations applicable with reasonable accuracy. For a linear behavior with particle fluence ($\Phi$) of both, $T_\textrm{c,i}(\Phi)=T_\textrm{c,p}+\partial T_\textrm{c}/\partial \Phi\cdot\Phi$ and $\rho_\textrm{n,i}(\Phi)=\rho_\textrm{n,p}+\partial\rho_\textrm{n}/\partial \Phi\cdot\Phi$,~\cite{Rullier2003} $\partial\rho_\textrm{n}/\partial T_\textrm{c}$ is constant and the relative change in $\alpha$ becomes: $\alpha/\alpha_\textrm{p}=(1-K_\rho(1-T_\textrm{c}/T_\textrm{c,p}))T_\textrm{c,p}/T_\textrm{c}$ with the experimentally accessible parameter $K_\rho=T_\textrm{c,p}/\rho_\textrm{n,p}\cdot\partial\rho_\textrm{n}/\partial T_\textrm{c}$. Note that the superfluid density changes inversely to $\alpha$ in the dirty limit since $\lambda_\textrm{p}^{2}/\lambda^{2}$ converges to $\alpha_\textrm{p}/\alpha$ for large $\alpha$. 
This prediction is compared with experimental data for YBCO films collected by Franz et al.~\cite{Franz1997} in Fig.~\ref{FigSuperfluid}. The data were obtained from films either irradiated with He ions or doped with Zn or Ni. Results on fast neutron irradiated films indicate $K_\rho=-16.5$, which was used to calculate the line graphs.  The agreement demonstrates that the prediction fits an universal suppression of the superfluid density with decreasing transition temperature, even with $K_\rho$ obtained from a system where the disorder was introduced differently. Note that using $\alpha_p=3$ instead of the dirty limit expression, reduces the slope at $T_\textrm{c}$, leading to a better agreement with the experimental data in the low disorder range relevant for fusion magnets.   

The ratio of the coherence length before and after irradiation, $\xi/\xi_\textrm{p}=\xi_\textrm{0}\sqrt{1+\alpha_\textrm{p}}/\xi_\textrm{0,p}\sqrt{1+\alpha}$ is needed for the calculation of the change of the depairing current density as well. This becomes $\sqrt{\xi_\textrm{0}l/\xi_\textrm{0,p} l_\textrm{p}}$ in the dirty limit, which can be expressed as $\sqrt{T_\textrm{c,p}\rho_\textrm{n,p}/T_\textrm{c}\rho_\textrm{n}}$ resulting in  $J_\textrm{dp}/J_\textrm{dp,p}=(T_\textrm{c}/T_\textrm{c,p})^{3/2}(\rho_\textrm{n,p}/\rho_\textrm{n})^{1/2}$.
\par\medskip
The reduction of the critical current due to flux creep, $A_\textrm{creep}=J_\textrm{c,exp}/J_\textrm{c}$ arises from a slow flux motion due to thermal activation processes with an average velocity, $v_\textrm{creep}$, generating an electric field $E=v_\textrm{creep}B$ perpendicular to the direction of the movement. $v_\textrm{creep}$ can be obtained from the hopping distance $a_\textrm{hop}$ and the frequency of thermal activation processes $\nu_\textrm{hop}=\nu_0\exp{(-U/k_\textrm{B}T)}$. A natural choice for the hopping distance in case of strong pinning is the distance between to neighboring vortices,  $a_\textrm{hop}=\sqrt{\Phi_0/B}.$ The so called attempt frequency was estimated to be $\nu_0=2.5\cdot10^7$\,/s nearly independent of temperature.~\cite{Maley1990} The activation barrier for depinning, $U$, is influenced by the vortex state, however, the power-law of the current-voltage characteristics ($E\propto J^n$), which is widely observed in experiments,~\cite{Thompson2008} requires a logarithmic energy barrier, $U=U_0\ln(J_\textrm{c}/J)$, with the power-law index $n$ just being $U_0/k_\textrm{B}T$.~\cite{Yeshurun1996} These relations lead to $E=\sqrt{\Phi_0B}\nu_0(J/J_\textrm{c})^n$ and with the electric field $E=E_\textrm{c}$ at $J=J_\textrm{c,exp}$ render $A_\textrm{creep}=(E_\textrm{c}/\sqrt{\Phi_0B}\nu_0)^{1/n}$.
\par
\begin{figure} 
\centering 
\includegraphics[, width = 0.8\columnwidth,clip]{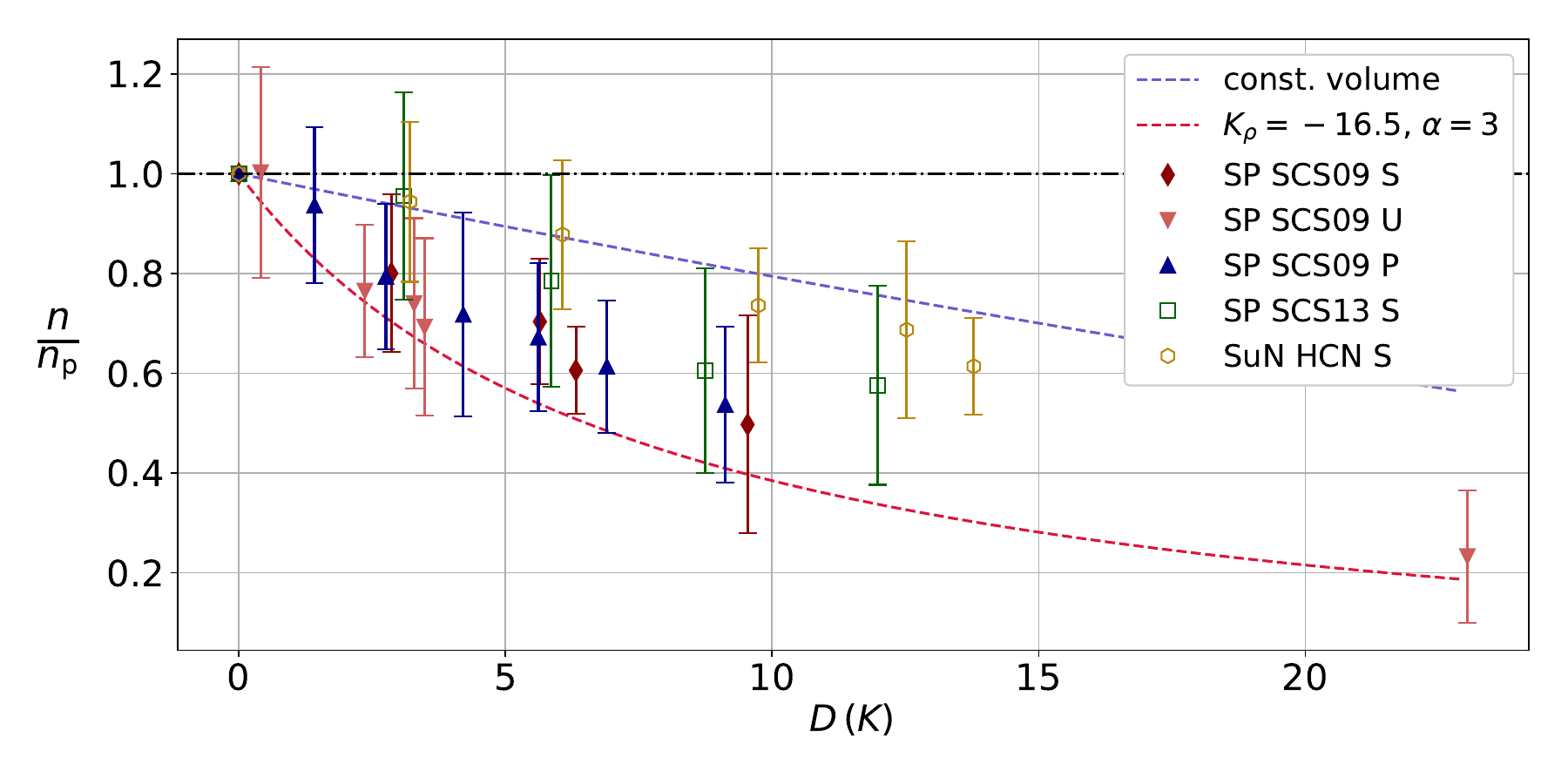} \caption{Relative change of the exponent $n$ of the power law $I\propto E^n$ in coated conductors at 30\,K, 15\,T. Simple models for the energy barrier against flux creep (line graphs) form an envelop around the experimental data; suggesting the decreasing superfluid density as the main driver for the change in $n$.} 
\label{Fign_norm}
\end{figure}

\medskip \newpage
Since $n$ is given by the activation barrier one can expect that it is closely related to the pinning energy gained in the volume to be depinned by thermal activation. The pinning energy is given by the condensation energy density ($E_\textrm{cond}\propto 1/\lambda^2\xi^2$) times an interacting volume, which is, in the simplest case of depinning from individual defects that are larger than the coherence length, the condensation energy of the core ($E_\textrm{cond}\xi^2\propto 1/\lambda^2$) times the size of the defect along the vortex core. In that case, $n$ decreases the same as the superfluid density, $n/n_\textrm{p}=\lambda_\textrm{p}^{2}/\lambda^{2}$. If the defects are smaller than $\xi$, $n$ decreases as the condensation energy density itself leading to $n/n_\textrm{p}=T_\textrm{c}^{2}/T_\textrm{c,p}^{2}$. The predicted behaviors are compared with experimental data in Fig.~\ref{Fign_norm}. Scattering of the experimental data is significant and the predictions for small and large defects form an envelope around the data. This suggests that the decrease in $n$ is driven by the enhanced scattering as well. A universal behavior of the change in $n$ is not to be expected anyway, since $U_0$ is related to but in general not the same as the pinning energy of a single defect and depends on details of the interaction of the vortex lattice with the defect landscape. Therefore, the decrease of the experimental $n$-value with disorder was fitted linearly for each tape as the input for the $A_\textrm{creep}$.

The degradation function is defined as
\begin{equation}
F_D(D):=\frac{J_\textrm{dp}(D)}{J_\textrm{dp,p}}\frac{A_\textrm{creep}(D)}{A_\textrm{creep,p}} =\sqrt{\frac{(\alpha_\textrm{p}+1)t_\textrm{c}^3}{\alpha_\textrm{p}(1-K_\rho(1-t_\textrm{c}))+t_\textrm{c}}}\left(\frac{E_\textrm{c}}{\sqrt{\Phi_0B}\nu_0}\right)^{1/n-1/n_\textrm{p}}
\label{EqDegrad}
\end{equation}
with $t_\textrm{c}:=T_\textrm{c}/T_\textrm{c,p}=1-D/T_\textrm{c,p}$. The exponent $1/n-1/n_\textrm{p}$ becomes $\partial n/\partial D \cdot D/n_\textrm{p}(1-\partial n/\partial D \cdot D)$ for the assumed linear change of $n=n_\textrm{p}(1-\partial n/\partial D \cdot D)$ with disorder
\par\medskip
All changes have been modeled on the basis of $\lambda$ and $\xi$ at 0\,K. The required ratios before and after introducing disorder should be the same at low temperatures, since the temperature dependence of superconducting mixed state parameters is given by power-laws of the form $(1-(T/T_\textrm{c})^a)^b$. Hence the observed changes in $T_\textrm{c}$ change the ratios insignificantly at low temperatures but become very important when approaching the transition temperature.

\section{Methods}
To study the degradation of high temperature superconductors (HTS) in radiation environments, Rare-Earth-Barium-Copper-Oxide (REBCO) based coated conductors were neutron or proton irradiated. \qty{27}{\milli\meter} long pieces were cut from commercial \qty{4}{\milli\meter} wide tapes supplied by SuperPower (SP) and SuNAM (SuN). For this study, we chose three different tapes, one from each supplier containing no artificial pinning centers (SuN HCN, SP SCS09) and one from SuperPower (SP SCS13) with \ce{BaZrO_3} nano-precipitates~\cite{Chen2009}. The superconducting layers of the samples provided by SP were deposited using metal-organic chemical vapour deposition (MOCVD). SuNAM exploits the reactive co-evaporation by deposition and reaction (RCE-DR) method for this purpose.
These samples were chosen to guarantee that the observed degradation behavior is independent of supplier (deposition method) and pinning landscape.

\begin{table}[h]
\renewcommand{\arraystretch}{1.2}
\centering
\begin{tabular}{ccccc}
identifier &  tape & material& dep. method \\ \cline{1-4}
\multicolumn{1}{c|}{$\;\;\;$ SP SCS09 $\;\;\;$ }&SCS4050 2009 & GdBCO&MOCVD\\

\multicolumn{1}{c|}{$\;\;\;$SP SCS13$\;\;\;$}&SCS4050 2013& (Y,Gd)BCO &MOCVD \\
\multicolumn{1}{c|}{$\;\;\;$SuN HCN$\;\;\;$} &HCN04150 & GdBCO &RCE-DR \\
\end{tabular}
\caption{Sample identifiers}\label{tab:samplelist}
\end{table}

All samples use a Hastelloy substrate with a textured \ce{MgO} layer made by IBAD (ion beam assisted deposition); the superconducting layer is \qty{1}{\micro\meter} and \qty{1.3}{\micro\meter} thick in the SuperPower and SuNAM tapes, respectively. The samples are coated with \qty{1}{\micro\meter} of \ce{Ag} and electrically stabilized by a \qty{\sim 20}{\micro\meter} copper layer. For proton irradiation, the copper layer was removed and the samples were bridged at a length of \qty{2}{\milli\meter} to a width of \qty{0.2}{\milli\meter} with a laser cutter. \par\medskip

The samples were measured applying a standard four probe measurement technique in the variable-temperature-insert (VTI) of a liquid helium cooled \qty{17}{\tesla} cryostat. $I_\mathrm{c}$ measurements were conducted at temperatures ranging from \qty{30}{\kelvin} to \qty{77}{\kelvin} and in fields up to \qty{15}{\tesla}, which was applied perpendicular to the the tape.
The power law $U=E_\mathrm{c}d(I/I_\mathrm{c})^n$ was fitted to the acquired data for determining the critical current and $n$-value with the electric field criterion $E_\mathrm{c} = \,$\qty{1}{\micro\volt\per\centi\meter} and the distance between the voltage taps $d$.
\par

We restrict our considerations to low temperatures, since fusion magnets are currently foreseen to operate at around 20\,K. Critical currents are very high at 20\,K in many tapes imposing thermal problems during measurements in our facilities, therefore, we choose 30\,K as a compromise. \par \medskip
 The critical temperature was determined by applying a \qty{10}{\milli\ampere} current to the sample, ramping from high to low temperatures at \qty{0.1}{\kelvin\per\minute} and measuring the voltage drop between the contact pins. The acquired voltage was derived with respect to the temperatur, rendering the maximum slope $k_\mathrm{max}$. The transition was defined as the temperature range where $k>0.6\,k_\mathrm{max}$ and fitted by a linear function.  The critical temperature was then obtained by intersecting the linear function with the $x$-axis. \par\medskip
Neutron irradiation was carried out in the TRIGA Mark II reactor at Atominstitut, TU Wien. The pre-characterized samples were welded in quartz-tubes and irradiated in the central irradiation facility with a fast neutron (\qty{>0.1}{\mega\electronvolt}) flux of $f_\mathrm{f}=3.5\times10^{16} \,\mathrm{m}^{-2}\mathrm{s}^{-1}$ at temperatures not exceeding \qty{70}{\celsius} up to a cumulative fluence of \qty{4.3e22}{\per\square\meter}.~\cite{Unterrainer2024} The neutron spectrum in the central irradiation facility of the TRIGA Mark II exhibits two peaks, one at high and one at thermal energies.~\cite{Weber1986} Low energy neutrons (\qty{<0.55}{\electronvolt}) are usually shielded with cadmium in irradiation experiments to approximate the neutron spectrum expected at the magnets in a fusion device.~\cite{Weber2011} In this study some sample sets were irradiated with shielding in place (S) and others without (U). This enables the introduction of vastly different defect size distribution by exploiting the fact that two Gd isotopes exhibit massive absorption cross sections for thermal neutrons. Upon absorbing a neutron, the nuclei enter an excited state, which decays by a gamma emission with a recoil energy of \qtyrange{29}{34}{\electronvolt},~\cite{Sickafus1992} just enough to displace the Gd atom from its lattice position, resulting in the introduction of point-like defects.~\cite{Unterrainer2024} On the other hand, if samples are shielded from low energy neutrons, the introduced defects are expected to resemble those expected in fusion. \par\medskip
Bridged samples were irradiated with \qty{1.2}{\mega\electronvolt} protons (P) at room temperature with a General Ionix \qty{1.7}{\mega\volt} tandem accelerator at the Plasma Science and Fusion Center of MIT.~\cite{Alexis2024} \qty{1.2}{\mega\electronvolt} protons introduce mainly small defects. High levels of disorder can be obtained with short irradiation times and avoiding problems with the generation of radio-isotopes; thus, the radiation tolerance of a superconductor can be tested with a comparably low effort.



\section{Appendix}

\begin{acknowledgments}
 This work has been carried out within the framework of the EUROfusion Consortium, funded by the European Union via the Euratom Research and Training Programme (Grant Agreement No 101052200 — EUROfusion). Views and opinions expressed are however those of the author(s) only and do not necessarily reflect those of the European Union or the European Commission. Neither the European Union nor the European Commission can be held responsible for them.\par\medskip
The authors would like to acknowledge D.X. Fischer and K.B. Woller from the Plasma Science and Fusion Center of MIT for providing the opportunity to proton irradiate one of the samples.
\end{acknowledgments}

\end{document}